\documentclass[aps, prb, preprint, superscriptaddress]{revtex4}
\usepackage{graphicx}
\usepackage{amsmath}
\usepackage{amssymb}

\newcommand*{\Sm}{SmFe$_3$(BO$_3$)$_4$}
\newcommand*{\Yb}{YbAl$_3$(BO$_3$)$_4$}
\newcommand*{\vect}[1]{\mathbf{#1}}

\begin{document}

\title{Sign change of polarization rotation under either time or space    inversion in magnetoelectric \Yb~}

\author{A. M. Kuzmenko}
\affiliation{Prokhorov General Physics Institute, Russian Academy of
Sciences, 119991 Moscow, Russia}
\author{V. Dziom}
\author{A. Shuvaev}
\author{Anna Pimenov}
\author{D. Szaller}
\affiliation{Institute of Solid State Physics, Vienna University of
Technology, 1040 Vienna, Austria}
\author{A. A. Mukhin}
\author{V. Yu. Ivanov}
\affiliation{Prokhorov General Physics Institute, Russian Academy of
Sciences, 119991 Moscow, Russia}
\author{A. Pimenov}
\affiliation{Institute of Solid State Physics, Vienna University of
Technology, 1040 Vienna, Austria}

\begin{abstract}

Materials with optical activity can rotate the polarization plane of transmitted light. The most typical example is the natural optical activity, which has the symmetry property of changing sign after space inversion but being invariant to time inversion. Faraday rotation exhibits the opposite: it is invariant to space inversion but changes sign after time reversal. Here, we demonstrate that in a magnetoelectric material, another type of polarization rotation is possible. This effect is investigated in magnetoelectric \Yb~under the viewpoint of time and space inversion symmetry arguments. We observe the sign change of the rotation sense under either time or space reversal. This investigation proves that the polarization rotation in \Yb~must be classified as gyrotropic birefringence, which has been discussed within the idea of time-reversal breaking in underdoped cuprates. The diagonal terms in the magnetoelectric susceptibility are responsible for the observed signal of gyrotropic birefringence. Further analysis of the experimental spectra reveals a substantial contribution of the natural optical activity to the polarization rotation. We also demonstrate that the observed activity originates from the magnetoelectric susceptibility.

\end{abstract}

\date{\today}


\maketitle

\section{Introduction}

One of the classical optical effects for linearly polarized waves is a rotation of the polarization plane during its propagation through a medium. The two best known effects of this type are natural optical activity and Faraday rotation~\cite{barron_book}, which are commonly called natural circular birefringence (NCB) and magnetic circular birefringence (MCB), respectively. The most important difference between these two effects is their relation to the fundamental inversions of space and time. Faraday rotation is not affected by the inversion of space, i.e., the simultaneous inversion of all coordinates but changes sign upon time reversal. Natural circular birefringence is transformed in the opposite manner: it changes sign upon space inversion and is not affected by time reversal. Of course, several optical effects, e.g., linear birefringence, are fully symmetrical in this sense: they are not affected by space or time inversion.

Approximately fifty years ago, it was realized that an additional type of polarization rotation must exist from the symmetry viewpoint. This effect is called gyrotropic birefringence and has the remarkable feature of changing sign for either time or space inversion~\cite{brown_jap_1963}. It has been demonstrated that gyrotropic birefringence should be ultimately connected to specific terms of the magnetoelectric susceptibility tensor~\cite{birss_phmag_1967,hornreich_pr_1968}. Alternatively, gyrotropic birefringence may be explained via the concept of spatial dispersion of the generalized dielectric permittivity~\cite{hornreich_pr_1968, pisarev_ferro_1994}. After its theoretical prediction, optical experiments~\cite{pisarev_pt_1991} and antiferromagnetic resonances~\cite{mukhin_jmmm_1988} in Cr$_2$O$_3$ were attributed to gyrotropic birefringence. Gyrotropic birefringence has also been investigated~\cite{kurumaji_prl_2017} in multiferroic (Fe,Zn)$_2$Mo$_3$O$_8$, where it was explained via a diagonal term of the magnetoelectric susceptibility, the so-called axion term.
However, in previous experiments, no formal symmetry proof was provided. Recently, the concept of gyrotropic birefringence has become highly topical again for explaining the results of time-reversal breaking in underdoped cuprate superconductors~\cite{varma_epl_2014}.

Multiferroics are materials with simultaneous electric and magnetic order\cite{smolenskii_ufn_1982, fiebig_jpd_2005, dong_advph_2015, tokura_rpp_2014, fiebig_natrev_2016}. The electric and magnetic counterparts are strongly coupled in these materials, which enables effective control of the electric polarization by a magnetic field and of the magnetization by an electric voltage. In the dynamic regime, the magnetoelectric coupling leads to the appearance of new magnetic modes with an electric excitation channel, which are termed electromagnons~\cite{pimenov_nphys_2006, aguilar_prl_2009, kida_josab_2009}. Cross-coupling of the electric and magnetic components in electromagnons has resulted in a wealth of strong optical effects, particularly at terahertz frequencies, such as directional dichroism~\cite{kezsmarki_prl_2011, takahashi_nphys_2012, takahashi_prl_2013, kezsmarki_nc_2014, bordacs_prb_2015, kezsmarki_prl_2015, toyoda_prl_2015, kuzmenko_prb_2015}, magnetochiral dichroism~\cite{bordacs_nphys_2012, kibayashi_nc_2014} or giant rotatory power~\cite{kuzmenko_prb_2014}.

Here, we experimentally demonstrate that magnetoelectric materials reveal one more fundamental optical effect with unusual symmetry properties. In the magnetoelectric \Yb, the polarization rotation reveals sign changes under time or space inversion operations. Experimentally, the time-inverted sample is realized by changing the direction of the magnetization, and the space inversion is modeled by rotating the sample, which simultaneously reverses the direction of the induced electric polarization.

Recent investigations of rare earth borates~\cite{vasiliev_ltp_2006, kadomtseva_ltp_2010}  have been initiated due to the discovery of the static electric polarization and multiferroic behavior in GdFe$_3$(BO$_4$)$_3$~\cite{zvezdin_jetpl_2005} and NdFe$_3$(BO$_4$)$_3$~\cite{zvezdin_jetpl_2006}. The magnetoelectric interactions in these materials have been successfully explained by the crystal field energy scheme of the rare earth ions and the level shifts in external electric and magnetic fields~\cite{zvezdin_jetpl_2005, zvezdin_jetpl_2006, popov_prb_2013}.  Later on, in samarium ferroborate \Sm, a giant magneto\emph{di}electric effect was observed and well reproduced by a dynamic extension of the static mechanism~\cite{mukhin_jetpl_2011}. The antiferromagnetic ordering of Fe$^{3+}$ ions in ferroborates and the interactions between iron and rare earth subsystems make the full description of the observed effects complicated. Although the crystal field levels of the rare earth subsystem play a decisive role, in theoretical models, ferroborates can be reduced to an iron subsystem with renormalized coupling constants. It may be expected that in borates without the magnetism of iron atoms, magnetoelectric effects purely from rare earths can be observed. Indeed, a quadratic magnetoelectric effect has been detected in TmAl$_3$(BO$_4$)$_3$~\cite{chaudhury_prb_2010} and HoAl$_3$(BO$_4$)$_3$~\cite{liang_prb_2011} with induced electric polarization that even exceeds static values in multiferroic ferroborates. In agreement with previous considerations, the magnetoelectricity in alumoborates can be accounted for by the crystal field splitting of the rare earth and the symmetry of the local environment~\cite{begunov_jetpl_2013, kadomtseva_prb_2014, ivanov_jetpl_2017}.

\subsection{Sign change of polarization rotation}

To understand the experimental results below, we begin with a simple theoretical consideration. Compared to samarium ferroborate, in calculating the light propagation, several contributions to the electrodynamic response cannot be neglected.  The polarization-plane rotation of electromagnetic radiation may be described in several equivalent manners~\cite{barron_book}. In the regime near visible light, an approach based on the spatial dispersion~\cite{agranovich_book, landau_book8} of the dielectric permittivity in the form of $\varepsilon_{i,k}(\omega, \mathbf{k})=\varepsilon(\omega,0)+i\gamma_{i,k,l} k_l$ is common. Here, only terms that are linear in $\mathbf{k}$ are included, and the effects that arise from the magnetic properties of the material are neglected. Particularly toward the terahertz frequency range, the approximation $k\rightarrow 0$ is reasonable, and the polarization rotation is described via the magnetoelectric susceptibilities~\cite{birss_phmag_1967, hornreich_pr_1968} $\chi^{me,em}$. However, simultaneously considering both approaches, the magnetoelectric and spatial dispersion effects enter additively into the final expressions, which suggests that both terms are equivalent and can be reduced to each other by a linear transformation~\cite{serdyukov_book}. To simplify the discussion, we illustrate the topic using the following  example.

Assuming light propagation along the $c$-axis (or $z$-axis), the relevant term in the picture of spatial dispersion is given by the off-diagonal elements of the dielectric permittivity $\varepsilon_{xy}=-\varepsilon_{xy}=ik\alpha$. Solving the Maxwell equations and neglecting all other susceptibilities results in the eigenmodes as circular waves with a refractive index of $ n_{12} = n_0 \pm \alpha$. This term leads to a polarization rotation during the propagation of linearly polarized waves by an angle that is proportional to $\alpha$. An alternative approach, which uses the magnetoelectric susceptibility term $\chi^{me}_{yy}$, results in a polarization rotation proportional to $\chi^{me}_{yy}$. However, in the general magnetoelectric case, the propagating eigenmodes are circularly polarized in the approximation $\chi^{me}_{yy}\ll 1$ only, and they are elliptical in the general case (see theory Section below).

\begin{figure}[tbp]
\begin{center}
\includegraphics[width=0.99\linewidth, clip]{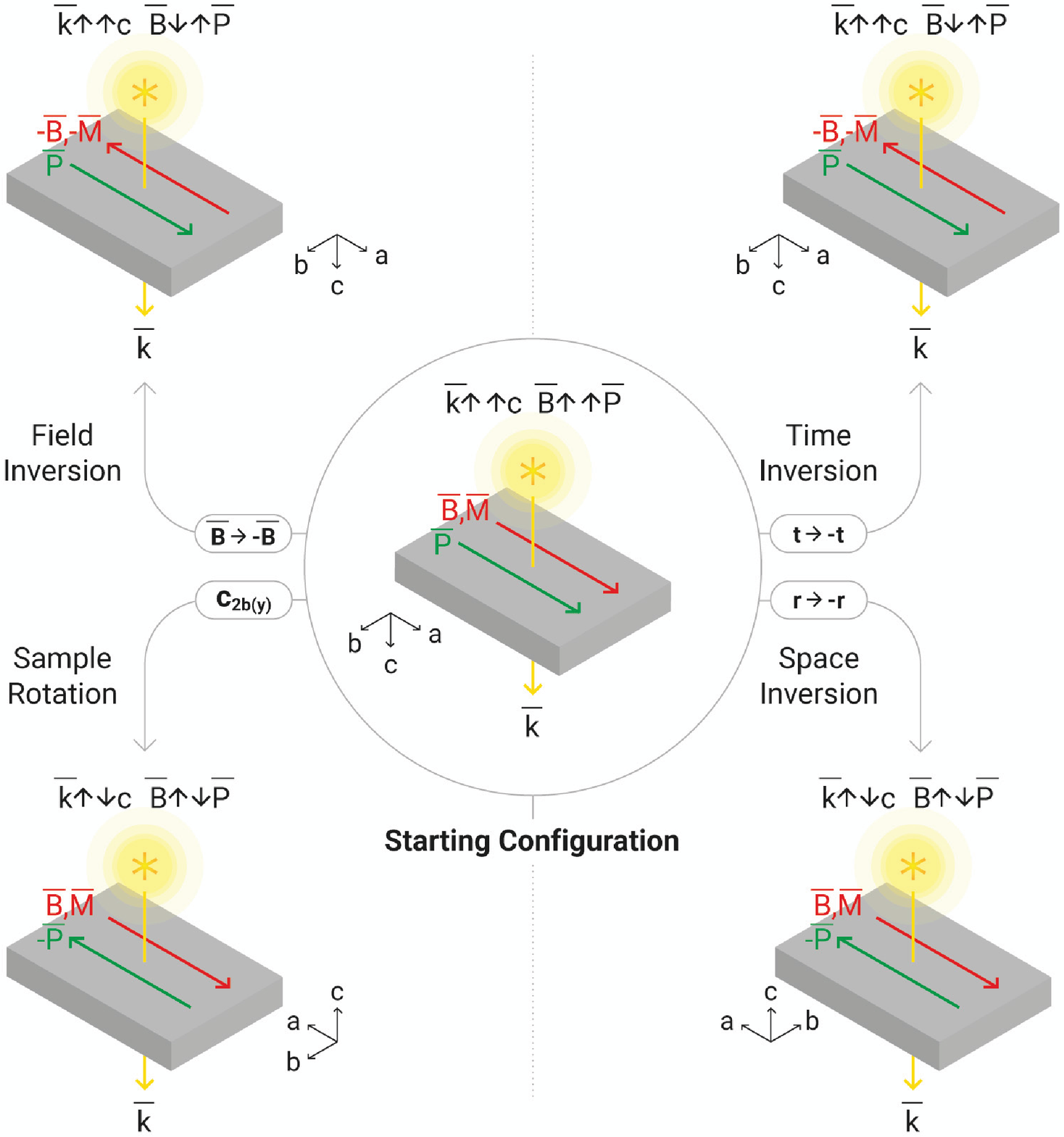}
\end{center}
\caption{\emph{Experimental geometry to prove the reversal of the polarization rotation by time and space inversions.} Middle: starting configuration with the indicated directions of relevant vectors and crystallographic axes. Top right: time inversion of the sample. Bottom right: space inversion of the sample. Left panels: experimental realization of the inversion symmetries as described in the text. The axes $a,b$, and $c$ correspond to the Cartesian axes $x,y$, and $z$, respectively. In \Yb~the external magnetic field $\bar{B}\|a$-axis induces a weak magnetization $M\|a$. In magnetoelectric alumo-borates and within the present geometry, the static electric polarization along the $a$-axis $P_a \sim B_a^2$ is induced~\cite{kadomtseva_prb_2014, kostyuchenko_ieee_2014, ivanov_jetpl_2017}, see  Eq.\,(\ref{eq_pol}).} \label{fig1}
\end{figure}

The symmetry of \Yb~enables us to prove the existence of gyrotropic birefringence in the terahertz transmission experiments. Here, the antisymmetric behavior can be investigated for both, time inversion and space inversion symmetries. The concept of the experimental arrangements is presented in Fig.~\ref{fig1}. We start with a reference experiment with the sample geometry given in the middle panel of Fig.~\ref{fig1}. In this case, $\bar{k}\upuparrows c$-axis, and $\bar{M}\upuparrows \bar{P}$. Here, $\bar{k}, \bar{M}$ and $\bar{P}$ are the wavevector, magnetization, and electric polarization, respectively.

To prove the unconventional character of the optical activity in \Yb, the symmetry properties must be investigated with respect to space and time symmetry operations. The effect of both symmetries is shown in the right panels of Fig.~\ref{fig1}. We recall that the space inversion reverses the direction of the electric polarization and of the $c$-axis (bottom right panel in Fig.~\ref{fig1}). Meanwhile, the time inversion only reverses the direction of the magnetization. The electric polarization and the direction of the $c$-axis are preserved for the time inversion.
In experiment, the inversion of external magnetic field properly simulate the time inversion as it reverses the direction of magnetic moments. Pure space inversion of the physical is not possible experimentally. However, in the case of \Yb the relevant vectors which determine the symmetry are $\bar{P}, \bar{M}$, and $\bar{k}$. As demonstrated in Fig.\,\ref{fig1} the space inversion in \Yb~ is properly modelled by $180^\circ$ sample rotation around the $b$-axis (or $c$-axis) with respect to starting configuration.

Figure\,\ref{fig1} shows two experimental possibilities to prove the symmetry of gyrotropic birefringence in \Yb~ experimentally: i) reversal of the external magnetic field and ii) $180^{\circ}$ rotation of the sample around the $b$-axis. A comparison of the left and right panels of Fig.~\ref{fig1} demonstrates that they represent the same experimental geometry for the symmetry conditions of magnetization, polarization and light propagation. Therefore, the set of two geometries in the left panels provides a good model to prove the effects of time and space inversions on \Yb. Strictly speaking, to realize space inversion for the optical activity, an enantiomorph crystal is necessary. However, the sign of the gyrotropic birefringence follows the static electric polarization. Therefore, the rotation of the sample around the $b$-axis provides sufficient data to model the corresponding symmetry condition.

\section{Experimental}

Large single crystals of \Yb~ with typical dimensions of $\sim 10 \times 10\times 1$
mm$^3$ were grown by crystallization from the melt on seed
crystals. A detailed analysis of the static magnetic and magnetoelectric properties of \Yb~can be found in Ref.~[\onlinecite{ivanov_jetpl_2017}]. The polarization rotation was investigated in the terahertz frequency range (40~GHz
$< \nu <$ 1000 GHz). Transmission experiments were conducted in a Mach-Zehnder
interferometer arrangement~\cite{volkov_infrared_1985}, which enabled us to
measure the amplitude and phase shift in a geometry with
controlled polarization of the radiation. In all experiments, linear incident polarization was used, and the polarization state of the transmitted wave was analyzed. The theoretical transmittance
curves \cite{shuvaev_sst_2012} for various geometries were
calculated from the susceptibilities within the Berreman
formalism \cite{berreman_josa_1972}. The experiments in external
magnetic fields up to 7~T were performed in a superconducting
split-coil magnet and within Voigt geometry with magnetic field perpendicular to the propagation of light. Due to possible misalignment of the sample with respect to magnetic field a small contribution of Faraday rotation can be present in the spectra. We estimate the absolute values of this effect as below 2 degree.

To investigate the polarization rotation in \Yb, several spectra in magnetic fields $\pm 6$\,T and $\pm 7$\,T as well as field dependencies at fixed frequencies of 70\,GHz 110\,GHz, and 120\,GHz were measured.
Below, basically the data at $\pm 6$\,T and 110\,GHz will be presented. All other results and fits showed qualitatively similar behavior.

\section{Theoretical considerations}

\subsection{Analysis of dynamic susceptibilities}

We begin this section with the symmetry analysis of the susceptibilities. The geometry, relevant in the present case, is given by $\mathbf{H}\|\mathbf{M}\|\mathbf{P}\|a$. Here $\mathbf{H}$ is the external magnetic field, $\mathbf{M}$ is the static magnetization, $\mathbf{P}$ is the  static electric polarization, and $a$ is the crystallographic axis  in the basis plane of the sample. We assume that the crystallographic $a,b,c$ axes are along cartesian $x,y,z$ coordinates. The symmetry of the problem corresponds to the point group $2_x$.

Dynamic magnetic and electric response of the system in the most general case may be written as:

\begin{equation}
\begin{array}{c}
\Delta \vect{m} = \chi^m \vect{h} + \chi^{me} \vect{e}, \\
\Delta \vect{p} = \chi^{em} \vect{h} + \chi^e \vect{e}. \\
\end{array}
\label{const_eqs}
\end{equation}

Within the point group $2_x$ the components of the susceptibility matrices may be written as:

\begin{equation}
\begin{array}{cc}
\hat{\chi}^m =
\left( \begin{array}{ccc}
\chi_{xx}^m & 0 & 0 \\
0 & \chi_{yy}^m & \chi_{yz}^m \\
0 & \chi_{zy}^m & \chi_{xx}^m \\
\end{array} \right) & \hat{\chi}^{me} =
\left( \begin{array}{ccc}
\chi_{xx}^{me} & 0 & 0 \\
0 & \chi_{yy}^{me} & \chi_{yz}^{me} \\
0 & \chi_{zy}^{me} & \chi_{zz}^{me} \\
\end{array} \right) \\[3em]
\hat{\chi}^{em} = \left( \begin{array}{ccc}
\chi_{xx}^{em} & 0 & 0 \\
0 & \chi_{yy}^{em} & \chi_{yz}^{em} \\
0 & \chi_{zy}^{em} & \chi_{zz}^{em} \\
\end{array} \right) & \hat{\chi}^e = \left( \begin{array}{ccc}
\chi_{xx}^e & 0 & 0 \\
0 & \chi_{yy}^e & \chi_{yz}^e \\
0 & \chi_{zy}^e & \chi_{zz}^e \\
\end{array} \right). \\
\end{array}
\label{suscept}
\end{equation}

All elements in Eq.~(\ref{suscept}) are in general case complex. If dissipation can be neglected, the following relation hold: $\chi_{ij}^m = (\chi_{ji}^m)^*$, $\chi_{ij}^e = (\chi_{ji}^e)^*$, and $\chi_{ij}^{me} = (\chi_{ji}^{em})^*$, where $()^*$ means complex conjugation.

In order to obtain the simplified susceptibility matrices we use the procedure similar to the case~\cite{kuzmenko_prb_2014, kuzmenko_prb_2015} of \Sm~ and write the thermodynamic potential of the system as function of magnetization and electric polarization:

\begin{eqnarray}\label{thermo}
\nonumber  && \Phi (\vect{m},\vect{P})= \\
\nonumber  &-& N[\mu_{\bot} (m_xH_x+m_yH_y)+\mu_{\|}H_zm_z]- \\
\nonumber &-& k_B TS(\vect{m})-P_x[E_x+c_2(H_ym_z+H_zm_y)+ \\
 &+& c_4(H_x m_x-H_ym_y)]- \\
\nonumber &-& P_y[E_y-c_2(H_x m_z+H_z m_x) -c_4(H_xm_y+H_ym_x)]- \\
\nonumber &-& P_zE_z+ \frac{1}{2\chi_{\bot}^E}(P_x^2+P_y^2)+\frac{1}{2\chi_{\|}^E}P_z^2 \ ,
\end{eqnarray}
where the first term represents the Zeeman energy of the ground  Yb$^{3+}$ ion doublet with magnetic moments $\mu_{\|}$ and $\mu_{\bot}$, along and perpendicular to the trigonal C3 axis, respectively, the second one is determined by  the entropy $S(\vect{m})$, the third and fourth terms represent the magnetoelectric coupling and the remaining tree terms give the electric part of the thermodynamic potential in external electric field. $N$ is the number of Yb$^{3+}$ ions.

Minimizing the free energy with respect to electric polarization we obtain:

\begin{eqnarray}
\nonumber P_x &=& \chi_{\bot}^E [E_x+c_2(H_ym_z+H_zm_y)+c_4(H_xm_x-H_ym_y)] \\
\nonumber   P_y &=& \chi_{\bot}^E [E_y-c_2(H_xm_z+H_zm_x)-c_4(H_xm_y+H_ym_x)] \\
\label{eq_pol} P_z &=& \chi_{\|}^E E_z
\end{eqnarray}

and

\begin{eqnarray*}
  M_x &=& -\frac{\partial \Phi}{\partial H_x} = N\mu_{\bot}m_x + c_4P_xm_x - P_y(c_2m_z+c_4m_y) \\
  M_y &=& -\frac{\partial \Phi}{\partial H_y} = N\mu_{\bot}m_y + P_x(c_2m_z-c_4m_y) - P_yc_4m_x \\
  M_z &=& -\frac{\partial \Phi}{\partial H_z} = N\mu_{\|}m_z + P_xc_2m_y - P_yc_2m_x \ .
\end{eqnarray*}

In the geometry with $H\|x\|a$ we get:

\begin{equation} \label{eq_pol2}
  m_x^0 \neq 0; \ P_x^0=\chi_{\bot}^E c_4 H_x m_x^0 \ ,
\end{equation}
where $m_x^0 \approx \tanh(\mu_{\bot} H_x/k_B T)$ is the static  value of the normalized magnetic moment of Yb$^{3+}$ ions along the  $x$-axis.

Similar to Ref.\,[\onlinecite{kuzmenko_prb_2014}] we solve the Landau-Lifshitz equations for dynamic magnetic and electric response. The susceptibility matrices are obtained in the approximation linear in coupling constants $c_2$ and $c_4$ as:

\begin{equation}
\begin{array}{cc}
\hat{\chi}^m =
\left( \begin{array}{ccc}
\chi_{xx}^m & 0 & 0 \\
0 & \chi_{yy}^m & \chi_{yz}^m \\
0 & \chi_{zy}^m & \chi_{xx}^m \\
\end{array} \right) & \hat{\chi}^{me} =
\left( \begin{array}{ccc}
\chi_{xx}^{me} & 0 & 0 \\
0 & \chi_{yy}^{me} & 0 \\
0 & \chi_{zy}^{me} & 0 \\
\end{array} \right) \\[3em]
\hat{\chi}^{em} = \left( \begin{array}{ccc}
\chi_{xx}^{em} & 0 & 0 \\
0 & \chi_{yy}^{em} & \chi_{yz}^{em} \\
0 & 0 & 0 \\
\end{array} \right) & \hat{\chi}^e = \left( \begin{array}{ccc}
\chi_{xx}^e & 0 & 0 \\
0 & \chi_{yy}^e & 0 \\
0 & 0 & \chi_{zz}^e \\
\end{array} \right) \ , \\
\end{array}
\label{suscept2}
\end{equation}
where the single components are given by:

\begin{eqnarray}\label{eq_comp}
\nonumber \chi_{xx}^m  &=& \chi^E_{\bot}c_4^2 (m_x^{0})^2  \\
\nonumber  \chi_{yy}^m &=& N\mu_{\bot} \frac{m_x^0}{H_x} R(\omega) \\
\nonumber  \chi_{yz}^m &=& \frac{i\omega}{\omega_0}N\mu_{\|}\frac{m_x^0}{H_x} R(\omega) \\
\nonumber  \chi_{zz}^m &=& \frac{\mu_{\|}}{\mu_{\bot}}N\mu_{\|}\frac{m_x^0}{H_x} R(\omega)  \\
  \chi_{xx}^{me} &=& (\chi_{xx}^{em})^* = \chi_{\bot}^E c_4 m_x^0 \\
\nonumber  \chi_{yy}^{me} &=& (\chi_{yy}^{em})^* = -\chi_{\bot}^E c_4 m_x^0(1+R(\omega)) - \frac{i\omega}{\omega_0} \chi_{\bot}^E c_2 m_x^0 R(\omega) \\
\nonumber  \chi_{zy}^{me} &=& (\chi_{yz}^{em})^*= \\
\nonumber  &=& \left[ \frac{i\omega}{\omega_0} \frac{\mu_{\|}}{\mu_{\bot}} \chi_{\bot}^E c_4 m_x^0 - (1+\frac{\mu_{\|}}{\mu_{\bot}}-\frac{\omega^2}{\omega_0^2})\chi_{\bot}^E c_2 m_x^0 \right] R(\omega) \\
\nonumber  \chi_{xx}^e &=& \chi^E_{\bot} \\
\nonumber  \chi_{yy}^e &=& \chi^E_{\bot} +  \frac{(\chi^E_{\bot})^2 m_x^0 H_x}{N \mu_{\bot}}(c_2^2+c_4^2)R(\omega)   \\
\nonumber  \chi_{zz}^e &=& \chi^E_{\|}
\end{eqnarray}
Here, $\chi_{\bot}^E$ is the background dielectric susceptibility, $m_x^0$ is the static magnetic polarization, $R(\omega)$ is the Lorentzian centered around $\omega_0 \propto H$, $c_4$ and $c_2$ are magnetoelectric constants that describe the chirality of \Yb, and  $R(\omega)=\omega_0^2/(\omega_0^2-\omega^2-i\omega g)$ is the resonance function with $\omega_0 =2 \mu_{\bot}H/\hbar$ and $g$ being the resonance frequency and linewidth, respectively.  Strictly speaking, the term $\chi_{xx}^m$ is quadratic in the coupling constants and may be neglected in the same approximation. On the contrary, the last term in $\chi_{yy}^e$ is the first nonzero term and it is resonant in frequency due to $R(\omega)$. Therefore, we keep this term in the present approximation.

\subsection{Electrodynamics of YbAl$ _{3}$(BO$ _{3}$)$ _{4}$ \label{sec_ed}}

 \subsubsection{General expressions}

We start with rewriting the materials relations in the form:

\begin{equation}
\begin{array}{c}
  \mathbf{B} = \hat{\mu }    \mathbf{H} +  \hat{\alpha }^{me}  \mathbf{E} \\
  \mathbf{D} = \hat{\alpha } ^{em}     \mathbf{H} + \hat{\varepsilon } \mathbf{E} \\
\end{array}
\end{equation}
Here $\hat{\mu }=\hat{1}+4\pi \hat{\chi }^{m} ,\  \hat{\varepsilon }=\hat{1}+4\pi \hat{\chi }^{e} ,\ \hat{\alpha }^{me,em} =4\pi \hat{\chi }^{me,em} $ and $\hat{\chi }^{m} ,\ \hat{\chi }^{e} ,\ \hat{\chi }^{me,em}$ are determined in the previous section.

The Maxwell equations $\nabla \times \mathbf{E} = - \mathbf{\dot{B}}/c , \nabla \times \mathbf{H} = \mathbf{\dot{D}}/c$ for the propagation vector $\mathbf{k} = \mathbf{n}\frac{\omega }{c} $ parallel to the $z$-axis
can be reduced to contain transverse components of the electric fields $E_{x,y}$ only :

\begin{equation}
\begin{array}{c}
  (n_{z}^{2} -n_{z10}^{2} )E_{x} -\delta _{1} n_{z} E_{y} = 0 \\
  \delta _{2} n_{z} E_{x} +{  (}n_{z}^{2} -n_{z20}^{2} )E_{y} = 0
\end{array}
\label{eq_exy}
\end{equation}
where

\begin{equation*}
  \begin{array}{c}
     n_{z10}^{2} = \varepsilon _{xx} \tilde{\mu }_{yy} -\chi _{xx}^{me} \chi _{xx}^{em} \tilde{\mu }_{yy}  /\mu _{xx} \\
     n_{z20}^{2} = \tilde{\varepsilon }_{yy} \mu _{xx} -\tilde{\chi }_{yy}^{em} \tilde{\chi }_{yy}^{me} \mu _{xx} /\tilde{\mu }_{yy}\\
     \delta _{1} =\tilde{\chi }_{yy}^{me} -\chi _{xx}^{em} \tilde{\mu }_{yy}/\mu _{xx} \\
     \delta _{2} =\chi _{xx}^{me} -\tilde{\chi }_{yy}^{em} \mu _{xx}^  /\tilde{\mu }_{yy} \\
\tilde{\mu }_{yy} =\mu _{yy} -\mu _{zy} \mu _{yz} /\mu _{zz} \\
\tilde{\varepsilon }_{yy} =\varepsilon _{yy} -\alpha _{zy}^{me} \alpha _{yz}^{em} /\mu _{zz}  \\
\tilde{\alpha }_{yy}^{me} =\alpha _{yy}^{me} -\mu _{yz} \alpha _{zy}^{me} /\mu _{zz}  \\
 \tilde{\alpha }_{yy}^{em} =\alpha _{yy}^{em} -\mu _{zy} \alpha _{yz}^{em} /\mu _{zz}
   \end{array}
\end{equation*}

The solution of Eq.\,(\ref{eq_exy}) is given by four elliptical eigenmodes with the eigenvalues of the refractive index being different for forward and backward directions:

\begin{equation}\label{eq_n12}
\begin{array}{c}
  n_{z1,2}^{2} ={\textstyle\frac{1}{2}} (n_{z10}^{2} +n_{z20}^{2} -\delta _{1} \delta _{2} )\pm \\
 \pm \sqrt{{\textstyle\frac{1}{4}} (n_{z10}^{2} +n_{z20}^{2} -\delta _{1} \delta _{2} )^{2} -n_{z10}^{2} n_{z20}^{2} } \ .
\end{array}
\end{equation}
The corresponding normalized eigenvectors of the modes are given by the relations:
\begin{equation}\label{eq_eigenvec}
  \begin{array}{c}
   {    e}_{{  1x}} {    }=\frac{n_{z1} \delta _{1} }{\sqrt{|n_{z1}^{2} -n_{z10}^{2} |^{2} +|n_{z1} \delta _{1} |^{2} } } \\
 {e}_{{  1y}} {    }=\frac{n_{z1}^{2} -n_{z10}^{2} }{\sqrt{|n_{z1}^{2} -n_{z10}^{2} |^{2} +|n_{z1} \delta _{1} |^{2} } }   \\
    {  e}_{{  2x}} {    }=\frac{n_{z2}^{2} -n_{z20}^{2} }{\sqrt{|n_{z2}^{2} -n_{z20}^{2} |^{2} +|n_{z2} \delta _{2} |^{2} } } \\
 {e}_{{  1y}} {    }=\frac{-n_{z2} \delta _{2} }{\sqrt{|n_{z2}^{2} -n_{z20}^{2} |^{2} +|n_{z2} \delta _{2} |^{2} } } \ .
  \end{array}
\end{equation}

As a result, the electromagnetic wave propagating in the media can be represented by the superposition of two eigenmodes
\begin{equation*}
  \vect{{  E}}{  (z,t)}=E^{(1)} \vect{e}_{1} \exp (i\omega t-ik_{1} z)+E^{(2)} \vect{e}_{2} \exp (i\omega t-ik_{2} z)
\end{equation*}
%
The electric field of the electromagnetic wave propagating for the distance z in the media is determined by

\begin{equation}\label{eq_prop}
  \vect{{  E}}{  (z)}=\hat{S}(z)\vect{E}_{0} \equiv \frac{1}{\Delta } \left(\begin{array}{cc} {e_{1x} } & {e_{2x} } \\ {e_{1y} } & {e_{2y} } \end{array}\right)\left(\begin{array}{cc} {e^{-ik_{1} z} } & {0} \\ {0} & {e^{-ik_{2} z} } \end{array}\right)\left(\begin{array}{cc} {e_{2y} } & {-e_{1y} } \\ {-e_{2x} } & {e_{1x} } \end{array}\right)\vect{E}_{0}
\end{equation}
%
where $\hat{S}(z)$ is the Jones matrix and $\Delta =e_{1x} e_{2y} -e_{1y} e_{2x} $.

 If the incident wave has a linear polarization $\vect{E}_{{  0}} =E_{o} (\cos \alpha ,\sin \alpha )$ the propagating wave is in general case elliptically polarized that can be characterized by the polarization plane rotation $\theta$ and ellipticity $\eta$:
%
\begin{equation}\label{eq_rot}
  {  tan(}\theta +{  i}\eta {  )}=\frac{E_{y} (z)}{E_{x} (z)} =\frac{(e_{1y} e_{2y} \cos \alpha -e_{2x} e_{1y} \sin \alpha )e^{-ik_{1} z} +(-e_{1y} e_{2y} \cos \alpha +e_{1x} e_{2y} \sin \alpha )e^{-ik_{2} z} }{(e_{1x} e_{2y} \cos \alpha -e_{2x} e_{1x} \sin \alpha )e^{-ik_{1} z} +(-e_{1y} e_{2x} \cos \alpha +e_{1x} e_{2x} \sin \alpha )e^{-ik_{2} z} }
\end{equation}

Neglecting the weak non-resonance terms in $\mu _{xx} ,\varepsilon _{yy} ,\alpha _{xx}^{em} ,\alpha _{xx}^{me} $  ($\chi _{xx}^{m} ,\chi _{yy}^{e} ,\chi _{xx}^{em} ,\chi _{xx}^{me} $) one can simplify the eigenvalues and eigenvectors as:

\begin{equation}\label{eq_n12sec}
  \begin{array}{c}
    n_{z1,2}^{2} ={\frac{1}{2}} (\varepsilon _{xx} \tilde{\mu }_{yy} +\tilde{\varepsilon }_{yy} \mu _{xx} ) \pm \\
\pm
 \sqrt{{\textstyle\frac{1}{4}} (\varepsilon _{xx} \tilde{\mu }_{yy} -\tilde{\varepsilon }_{yy} \mu _{xx} )^{2} +\varepsilon _{xx} \mu _{xx} \tilde{\alpha }_{yy}^{em} \tilde{\alpha }_{yy}^{me} } = \\
=\varepsilon _{\bot }^{0} + 1/2(\varepsilon _{\bot }^{0} \Delta \mu +\Delta \varepsilon _{c4} +\Delta \varepsilon _{c2} )R(\omega) \pm \\
\pm
 R(\omega)\sqrt{1/4 (\varepsilon _{\bot }^{0} \Delta \mu +\Delta \varepsilon _{c4} +\Delta \varepsilon _{c2} )^{2} +\varepsilon _{\bot }^{0} \Delta \mu \Delta \varepsilon _{c2} (\omega ^{2} /\omega _{0}^{2} -1)} \ ,

  \end{array}
\end{equation}
where

\begin{equation}\label{eq_eyysec}
\begin{array}{l}
\mu _{xx} \approx {\rm 1},{\rm               }\varepsilon _{xx} \approx \varepsilon _{\bot }^{0}\ , \alpha _{xx}^{em} =\alpha _{xx}^{me} \approx 0 \\
\tilde{\mu }_{yy} =\mu _{yy} -\mu _{zy} \mu _{yz} /\mu _{zz} \approx \mu _{yy} =1+\Delta \mu R(\omega ) \\
 \tilde{\varepsilon }_{yy} =\varepsilon _{yy} -\alpha _{zy}^{me} \alpha _{yz}^{em} /\mu _{zz} \approx \varepsilon _{yy} =\varepsilon _{\bot }^{0} +(\Delta \varepsilon _{c2} +\Delta \varepsilon _{c4} )R(\omega ) \\
 \tilde{\alpha }_{yy}^{me} =\alpha _{yy}^{me} -\mu _{yz} \alpha _{zy}^{me} /\mu _{zz} \approx \alpha _{yy}^{me} =\left(-\sqrt{\Delta \mu \Delta \varepsilon _{c4} } +i\frac{\omega }{\omega _{0} } \sqrt{\Delta \mu \Delta \varepsilon _{c2} } \right)R(\omega ) \\
 \tilde{\alpha }_{yy}^{em} =\alpha _{yy}^{em} -\mu _{zy} \alpha _{yz}^{em} /\mu _{zz} \approx \varepsilon _{yy} =\left(-\sqrt{\Delta \mu \Delta \varepsilon _{c4} } -i\frac{\omega }{\omega _{0} } \sqrt{\Delta \mu \Delta \varepsilon _{c2} } \right)R(\omega )\\
\Delta\mu=4\pi N \mu_{\bot} m_x^0 / H_x \\
\Delta \varepsilon_{c2,4}=4\pi (\chi_{\bot}^E c_{2,4})^2 m_x^0 H_x/N/\mu_{\bot}

\end{array}
\end{equation}


\begin{figure}[tbp]
\centering{}\includegraphics[width=0.5\linewidth,clip]{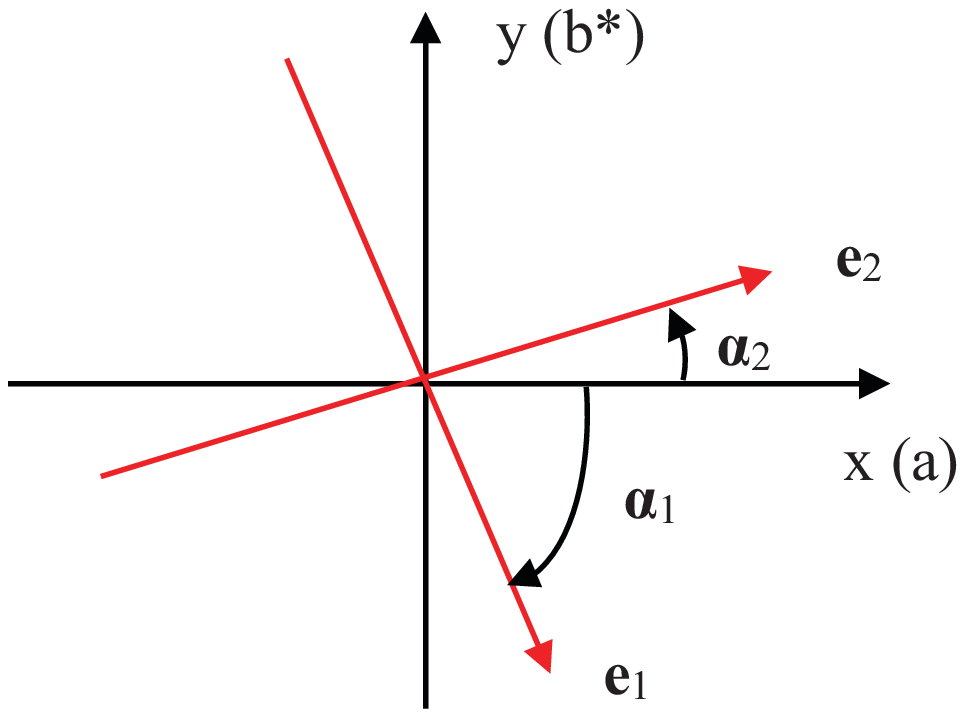}
\caption{\emph{Orientation of the eigenvectors of two eigenmodes in case of gyrotropic birefringence.} In this approximation the optical activity and dissipation are neglected, see Eqs.\,(\ref{eq_simpl_e},\ref{eq_angl}). The two modes become orthogonal far from the resonance $|\omega _{0} -\omega |\gg \omega _{0} $ when $\rho_{1\approx } - \rho_{2}$.}
     \label{fig_th}
\end{figure}

\subsubsection{Gyrotropic birefringence}


To consider the pure effect of the gyrotropic birefringence, we neglect the terms with optical activity, i.e. put $\Delta \varepsilon _{c2} =0$. In this case the eigenvalues and eigenvectors are determined by gyrotropic birefringence associated with  $\chi _{yy}^{me} \sim \sqrt{\Delta \mu \Delta \varepsilon _{c4} } $:

\begin{equation}\label{eq_simpln}
  \begin{array}{c}
     {{\rm   }n_{z1}^{{\rm 2}} =\varepsilon _{\bot }^{0} +(\varepsilon _{\bot }^{0} \Delta \mu +\Delta \varepsilon _{c4} )R(\omega ),} \\

{{\rm   }n_{z2}^{{\rm 2}} =\varepsilon _{\bot }^{0} }
   \end{array}
\end{equation}

\begin{equation}\label{eq_simpl_e}
  \begin{array}{c}
     {\rm     }\frac{{\rm e}_{{\rm 1x}} }{{\rm e}_{{\rm 1y}} } =\rho _{1} =-n_{z1} \sqrt{\Delta \mu /\Delta \varepsilon _{c4} } = \\ =-\sqrt{(\varepsilon _{\bot }^{0} +(\varepsilon _{\bot }^{0} \Delta \mu +\Delta \varepsilon _{c4} )R(\omega ))\Delta \mu /\Delta \varepsilon _{c4} } \approx \\ \approx -\sqrt{\varepsilon _{\bot }^{0} \Delta \mu /\Delta \varepsilon _{c4} }  \\

 \frac{{\rm e}_{{\rm 2y}} }{{\rm e}_{{\rm 2x}} } =\rho _{2} =\sqrt{\varepsilon _{\bot }^{0} \Delta \mu /\Delta \varepsilon _{c4} }
   \end{array}
\end{equation}

In the non-dissipative media the eigenvectors (i.e. $\rho_{1,2}$) are real and thus the corresponding eigenmodes  are linearly polarized but they are not orthogonal in the general case. Their orientation is determined by

\begin{equation}\label{eq_angl}
  \begin{array}{c}
     {\rm     ctan}\alpha _{{\rm 1}} =\frac{{\rm e}_{{\rm 1x}} }{{\rm e}_{{\rm 1y}} } =\rho _{1} =-n_{z1} \sqrt{\Delta \mu /\Delta \varepsilon _{c4} } = \\
=-\sqrt{(\varepsilon _{\bot }^{0} +(\varepsilon _{\bot }^{0} \Delta \mu +\Delta \varepsilon _{c4} )R(\omega ))\Delta \mu /\Delta \varepsilon _{c4} } \approx \\ \approx -\sqrt{\varepsilon _{\bot }^{0} \Delta \mu /\Delta \varepsilon _{c4} }  \\

{\rm tan}\alpha _{{\rm 2}} =\frac{{\rm e}_{{\rm 2y}} }{{\rm e}_{{\rm 2x}} } =\rho _{2} =\sqrt{\varepsilon _{\bot }^{0} \Delta \mu /\Delta \varepsilon _{c4} }
   \end{array}
\end{equation}

If the incident linear polarization coincides with one of the eigenmodes, e.g., $\vect{E}_{0} =E_{0} (\cos \alpha _{1} , \sin \alpha _{1} )$, the electromagnetic wave corresponding to $n_{z1}$ remains linearly polarized and propagates without a polarization rotation (i.e. $\theta$=0). In this case the second mode with $n_{z2}$ is not excited. Similarly, for an incident wave polarized as  $\vect{E}_{{0}} =E_{0} (\cos \alpha _{2} ,{\rm     }\sin \alpha _{2} )$ only the mode with $n_{z2}$ propagates without polarization rotation while the mode n${}_{z1}$ is not excited (Fig.\,\ref{fig_th}).

Beyond the resonance region $|\omega _{0} -\omega |\gg \omega _{0} $ when $\rho_1 \approx  - \rho_{2}$, both eigenmodes become orthogonal ($\alpha _{{2}} -\alpha _{1} \approx \pi /2$):
\begin{equation*}
  {\rm     ctan}\alpha _1 \approx -{\rm tan}\alpha _2 =\rho _{2} =\sqrt{\varepsilon _{\bot }^{0} \Delta \mu /\Delta \varepsilon _{c4}} \ .
\end{equation*}
Note that the orientation of the eigenmodes with respect to the crystal axes is determined by $\rho _{2} =\sqrt{\varepsilon _{\bot }^{0} \Delta \mu /\Delta \varepsilon _{c4} } $ that can vary in a wide range.

\subsubsection{Jones (gyrotropic) birefringence}

The considerations above take into account not only the magnetoelectric coupling but also the anisotropy of permittivity and permeability. In the last step let us neglect this anisotropy, quadratic magnetoelectric contributions, and assume that $\varepsilon _{xx} =\tilde{\varepsilon }_{yy} =\varepsilon _{\bot }^{0} ,\mu _{xx} =\tilde{\mu }_{yy} =1$. In this case  Eq.\,(\ref{eq_simpln}) for the refractive index in the forward direction can be further simplified to

\begin{equation} \label{eq_nsmpl}
  n_{z1,2} = \sqrt{n_{z0}^{2} + 1/4\delta_0^2} \pm 1/2 \delta_0 \approx n_{z0}\pm 1/2 \delta_0,
\end{equation}

where $n_{z0}^{2} =\varepsilon _{\bot }^{0} ,\delta _{0} =\chi _{yy}^{me} -\chi _{xx}^{em} \ll \varepsilon _{\bot }^{0} $.

The corresponding eigenvectors are given by

\begin{equation}\label{eq_smpl_e2}
  \left(\frac{{ E}_{{x}} }{{ E}_{{ y}} } \right)_{{ 1,2}} =\frac{n_{z1,2} \delta _{0} }{n_{z1,2}^{2} -n_{z0}^{2} } \approx \mp {1} \ .
\end{equation}

Therefore, these modes are linearly polarized and their main optical axes are aligned by $\mathrm{\pm}$45$^{o}$ with respect to the a-axis and for $H\|a$-axis.

This type of birefringence was predicted by Jones~\cite{jones_josa_1948}. It has been called Jones birefringence and was first experimentally observed~\cite{roth_prl_2000} in liquids in parallel electric and magnetic fields.

\begin{figure}[tbp]
\begin{center}
\includegraphics[width=0.8\linewidth, clip]{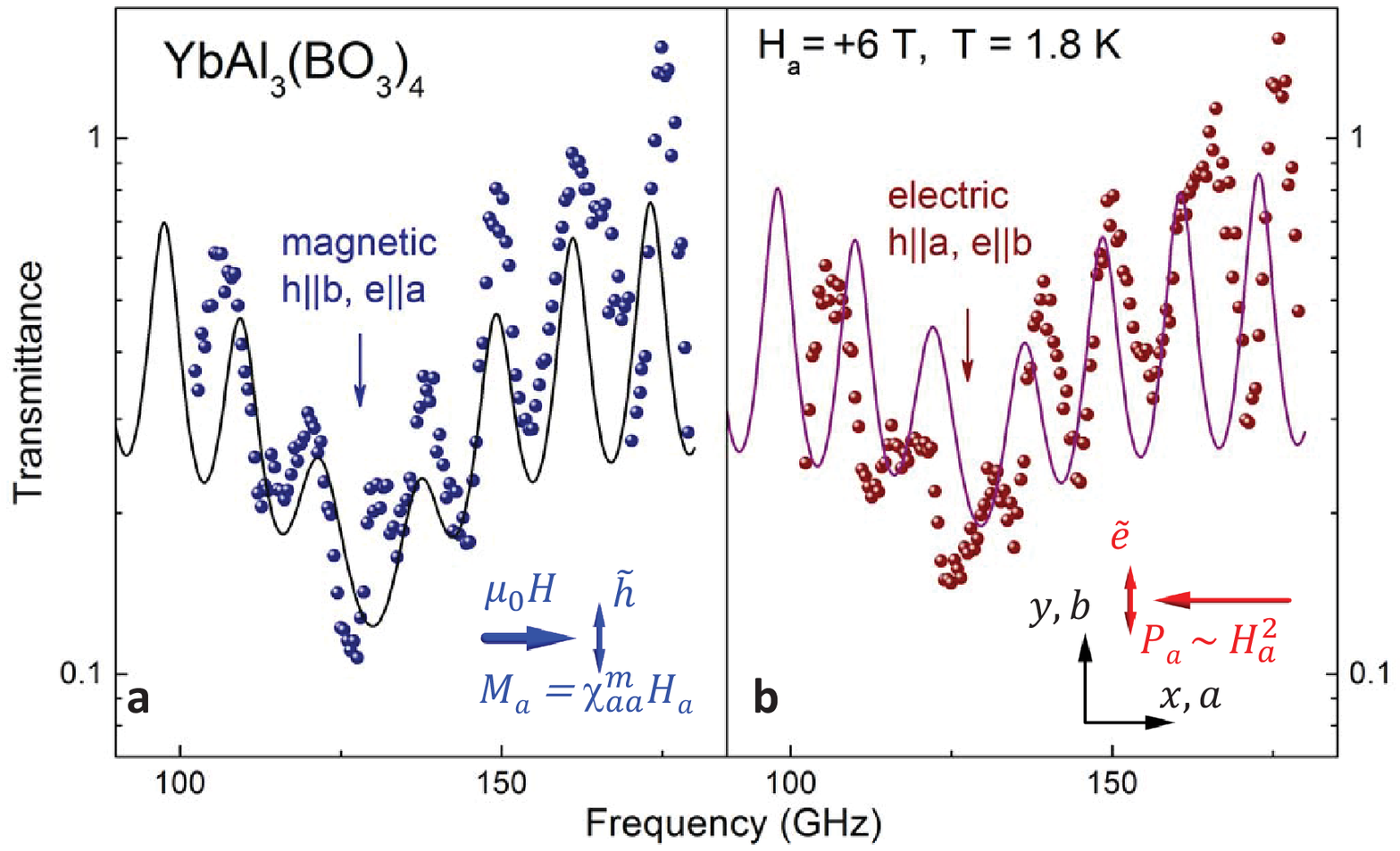}
\end{center}
\caption{\emph{Excitation of the electromagnon in \Yb.} \textbf{a}. Transmission spectra of the \Yb~sample in the geometry with a magnetic excitation channel, $h\|b, e\|a$. Blue symbols represent the data for the parallel orientation of the polarizer and analyzer. The arrow indicates the position of the electromagnon for $B_a = +6$~T. \textbf{b}. The spectra in the geometry with an electric excitation channel, $h\|a, e\|b$. Pictograms schematically show the excitation condition by the \emph{ac} components of electric and magnetic fields with respect to static vectors $M, P$ and $H$. The periodic oscillations in the spectra are due to Fabry-P\'{e}rot resonances on the sample surfaces. } \label{ftran0}
\end{figure}

\section{Results and Discussion}

Figure\,\ref{ftran0} shows the typical frequency-dependent transmission of \Yb~for two characteristic geometries with the light propagating along the c-axis: $h\|a, e\|b$ and $h\|b, e\|a$. The external magnetic field $H\|a$-axis induces a weak magnetization $M\|a$. In magnetoelectric alumoborates and within the present geometry, the static electric polarization along the $a$-axis $P_a \sim H_a m_0(H_a) \sim H_a^2$ is induced according to Eq.\,(\ref{eq_pol}), see also Refs.\,[\onlinecite{kadomtseva_prb_2014, kostyuchenko_ieee_2014, ivanov_jetpl_2017}]. In addition, similar to magnetoelectric ferroborates, a new electromagnon mode appears in the millimeter-wave spectra, showing electric, magnetic, and magnetoelectric activities~\cite{kuzmenko_jetp_2011, kuzmenko_prb_2014, kuzmenko_prb_2015, mukhin_ufn_2015}. In a simple approximation, the resonance frequency of the electromagnon in \Yb~is proportional to external magnetic field (see Fig.~\ref{figfld}\textbf{c} below), and the excitation conditions by the $ac$ fields are schematically shown in Fig.~\ref{ftran0}. Physically, the electric dipole activity of the electromagnon originates from $c_2$ and $c_4$ terms in the free energy, Eq.\,(\ref{thermo}), leading to $\Delta \varepsilon_{c2,4}$ terms in electric and magnetic susceptibilities Eqs.\,(\ref{eq_comp},\ref{eq_n12sec},\ref{eq_eyysec}).

The excitation conditions and transmission spectra in Fig.~\ref{ftran0}, as it was demonstrated~\cite{kuzmenko_prb_2014} for \Sm, show that
the electromagnon mode may be selectively excited by either an electric or magnetic component of the terahertz radiation. However, because of the magnetoelectric character of the mode, both excitation conditions are coupled, leading to the polarization rotation of light propagating inside the sample. The rotation in \Yb~is more complex: the observed effect can be identified as gyrotropic (or nonreciprocal) birefringence because it changes sign upon undergoing both space and time inversion symmetry operations. In addition, the observed asymmetry of the spectra demonstrates the existence of natural polarization rotation in \Yb. Importantly, this rotation can be attributed to the dynamic magnetoelectric effect.

\begin{figure}[tbp]
\begin{center}
\includegraphics[width=0.99\linewidth, clip]{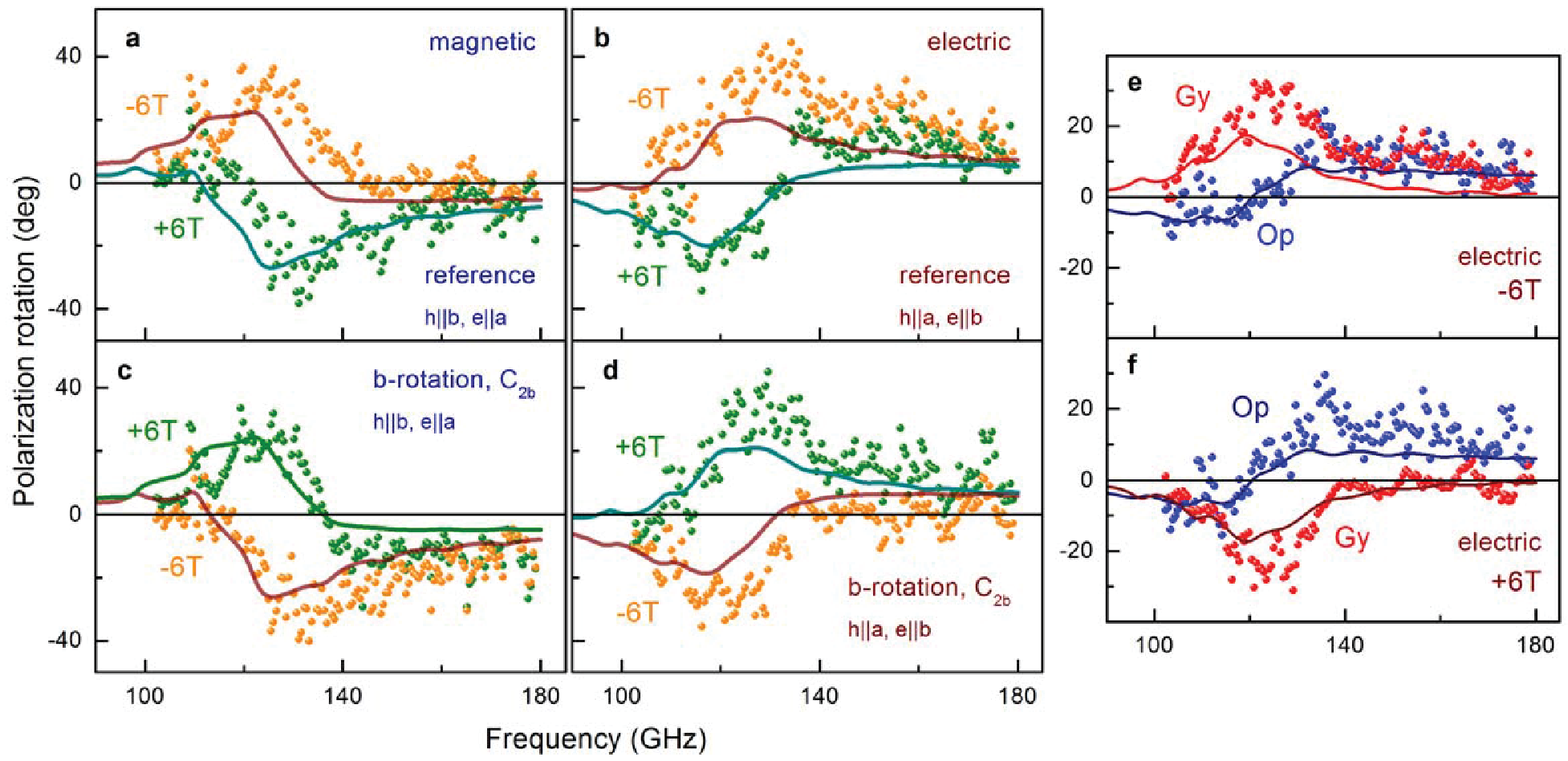}
\end{center}
\caption{\emph{Experimental test of the gyrotropic birefringence in \Yb.} \textbf{a,c}: magnetic excitation channel; \textbf{b,d}: electric channel. Panels \textbf{a,b} show the result in the reference geometry (middle part of Fig.~\ref{fig1}). The results in \textbf{c,d} are measured for the sample rotated by $180^{\circ}$, which simulates space inversion (\textit{cf}. Fig.~\ref{fig1}).  Symbols denote the experiment results, and solid  lines represent the results of the theoretical model according to Eq.\,(\ref{eq_rot}). \textbf{e,f}: Estimation of the relative effects from the gyrotropic birefringence (Gy) and optical activity (Op) in the geometry within the electric excitation channel.
} \label{ffreq}
\end{figure}

Static electric polarization in alumoborates is basically described by the term $P_a \sim H_a m_a - H_b m_b \sim H_a^2-H_b^2$, where the external magnetic field is aligned within the $ab$ plane (see Eq.\,(\ref{eq_pol}) and Refs.\,[\onlinecite{kadomtseva_prb_2014, kostyuchenko_ieee_2014, ivanov_jetpl_2017}]). This expression can be obtained from symmetry arguments~\cite{zvezdin_jetpl_2005, zvezdin_jetpl_2006} for the trigonal symmetry within the R32 space group. Here, we only reproduce the relevant terms for our experiments. We recall that in the case of alumoborates and specifically of \Yb, no spontaneous electric polarization exists in zero magnetic field.
The pictograms in Fig.~\ref{ftran0} illustrate the relevant geometry for the present experiment with $H\|a, P_a < 0$. In this configuration, the electromagnon may be selectively excited by a linear polarization via $h\|b$ (magnetic channel, $\mu_{yy}$) or $e\|b$ (electric channel, $\varepsilon_{yy}$). In both cases the rotation of the polarisation plane is due to the magnetoelectric term $\alpha_{yy}^{me}$.

As described in the theory Section, the eigenmodes in \Yb~ are elliptical waves $(E_x,E_y)$, where $E_x$ and $E_y$ are the transverse components of the $ac$ electric field. The ratio that determines the sign of the polarization rotation is $(E_x/E_y)$ is determined by $ \chi_{yy}^{em}$, Eq.\,(\ref{eq_comp}):
\begin{equation}\label{chiyy}
  \chi_{yy}^{me} =  -\chi_{\bot}^E c_4 m_x^0(1+R(\omega)) - \frac{i\omega}{\omega_0} \chi_{\bot}^E c_2 m_x^0 R(\omega) \ ,
\end{equation}
We recall that $c_4$ and $c_2$ are material constants that describe the chirality of \Yb.
The first term in Eq.~(\ref{chiyy}) is responsible for the gyrotropic birefringence and is odd with respect to both space and time inversions. The second term  describes the natural optical activity and is only reversed after space inversion. This behavior may be observed from the following arguments. The coefficients $c_4$ and $c_2$ in Eq.~(\ref{chiyy}) represent the chirality of the crystal structure. They change sign after space inversion (i.e., for another enantiomorph structure) and are not sensitive to time inversion. Meanwhile, the magnetization $m_0$ and magnetic resonance frequency $\omega_0$ do not change after space inversion and are reversed by time inversion. Because coefficients $c_4$ and $m_0$ are simultaneously present in the gyrotropic birefringence, the latter changes sign after both time and space inversions. The optical activity term contains $c_2$,  $m_0$, and $w_0$. Therefore, the effects of the time inversions cancel, and only the space inversion due to $c_4$ remains.

\begin{figure}[tbp]
\begin{center}
\includegraphics[width=0.6\linewidth, clip]{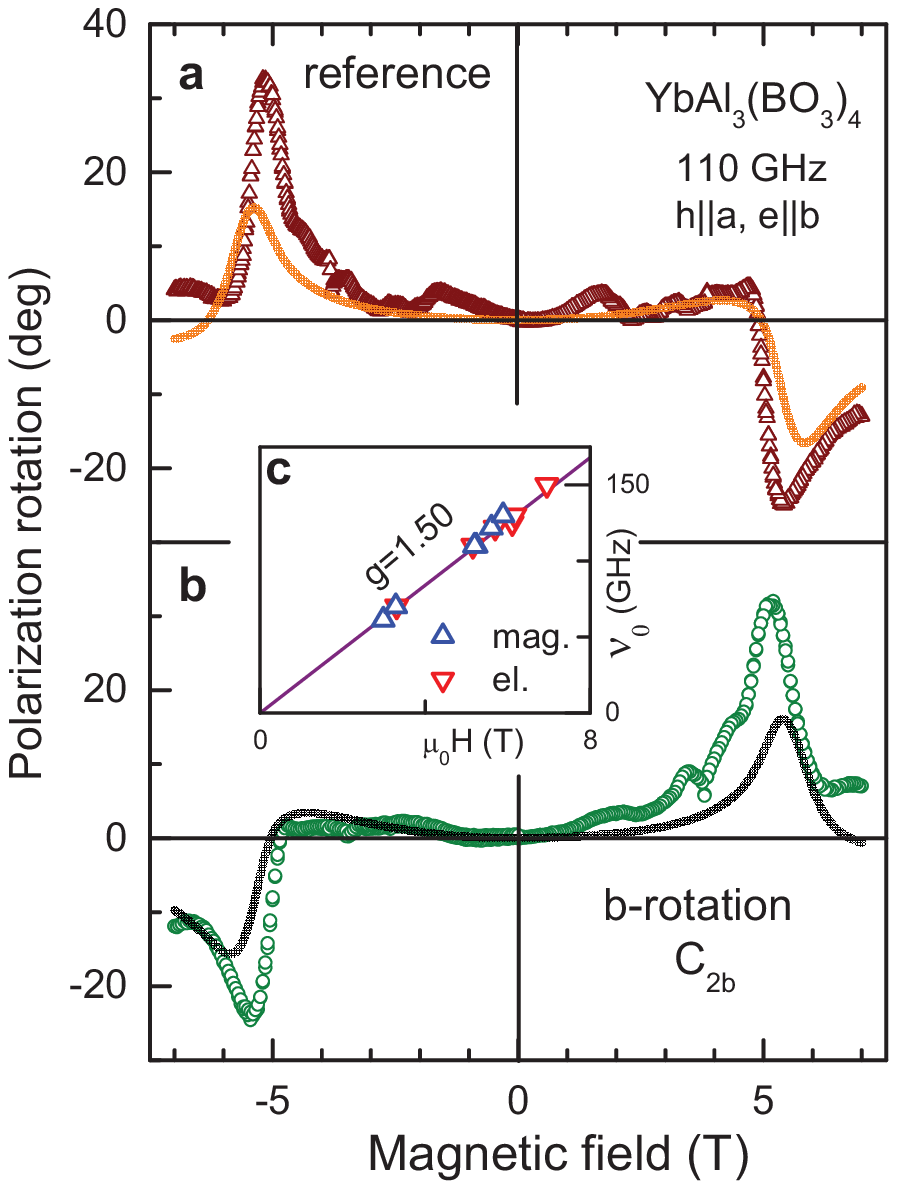}
\end{center}
\caption{\emph{Magnetic field sweeps in \Yb.} \textbf{a}: Reference geometry for the positive and negative directions of the magnetic field. The reversal of the signal as a function of magnetic field demonstrates the time asymmetry of the effect. \textbf{b}: The same experiments for the sample rotated around the b-axis. The sign of the polarization rotation is inverted compared to the reference geometry, which proves the space asymmetry.  Symbols denote the experimental results, and the solid line denotes the results of the theoretical model, which includes the gyrotropic birefringence and natural optical activity, with the parameters fixed by the data in Fig.~\ref{ffreq}. \textbf{c}: Magnetic field dependence of the electromagnon frequency. } \label{figfld}
\end{figure}

Figure\,\ref{ffreq} shows the experimental results of the polarization rotation according to the arguments above. The experimental geometries are given in the left panels of Fig.~\ref{fig1} and are equivalent to either time inversion of the starting geometry (reversal of the magnetic field) or space inversion (rotation of the sample around the $b$-axis).
In Fig.~\ref{ffreq}, all four possible geometries are compared on the same scale to demonstrate the inversion of the rotation signal in both time and space inversion symmetries. In all geometries, the rotation angle near the resonance is inverted if the external magnetic field changes sign. Importantly, the sequence of rotations also reverses after the sample rotates around the $b$-axis, which proves the expected reversal of the signal with respect to the space inversion. We note that compared to the spectra in Fig.~\ref{ftran0}, the Fabry-P\'{e}rot oscillations on the surfaces of the sample are not observed because of the suppression of the polarization rotation after internal reflections.

Concerning the rotation around the electromagnon mode near $\pm 6$~T, the data reveal an approximate antisymmetric behavior in the magnetic field. That is, the rotation direction changes after reversing the magnetic field, which qualitatively resembles the conventional experiments with Faraday rotation~\cite{zvezdin_book}. However, we recall that our experiment has been conducted in the Voigt geometry, i.e., $\bar{k} \bot H$, where the classical Faraday signal is zero.

In addition,  a non-perfect asymmetry is observed after the field reversal in Fig.~\ref{ffreq}, and the data cannot be adequately described by the model with only gyrotropic birefringence. The additional contribution is clearly not sensitive to the reversal of the magnetic field and to the sample rotation but changes sign between two excitation geometries. This second contribution corresponds to the natural optical activity and is generated by the second term in the magnetoelectric susceptibility  $\chi_{yy}^{me}$ in Eq.~(\ref{chiyy}). The effects from gyrotropic birefringence and from the optical activity are of comparable amplitudes in \Yb. To illustrate this schematically we may utilize the approximation of thin sample in which the rotation is simply proportional to the magnetoelectric susceptibility, and we neglect all other contributions. In this approximation the optical activity may be roughly obtained, e.g., as a sum of spectra in reference geometry and after $\mathrm{C_{2b}}$ rotation, respectively. Similarly, the difference of both spectra gives an estimate for the gyrotropic birefringence. The results are given in panels \textbf{e,f}. Within the notations of Fig.~\ref{ffreq}\,\textbf{e,f} the calculation procedure may be written as
$\mathrm{Op} = (\theta(\mathrm{ref})+\theta(\mathrm{C_{2b}}))/2$ and $\mathrm{Gy} = (\theta(\mathrm{ref})-\theta(\mathrm{C_{2b}}))/2$.
The solid lines in panels \textbf{e,f} give the exact contributions of both rotation angles according to Eq.\,(\ref{eq_rot}) by taking into account one term in Eq.\,(\ref{chiyy}) only.

In addition, a series of experiment in a sweeping magnetic field has been performed. These results are given in Fig.~\ref{figfld}. The results in the reference geometry are shown in panel \textbf{a}. The resonance rotation signal clearly changes sign with the reversal of the magnetic field. In agreement with the spectra in Fig.~\ref{ffreq}, this result demonstrates the time inversion asymmetry of the gyrotropic birefringence. The effect is inverted again after rotating the sample around the $b$-axis, thus proving the space inversion asymmetry. The theoretical curves in Fig.~\ref{figfld} are given by the solid lines and are qualitatively consistent with the experiment. Because all model parameters have been fixed by the fits of the transmission spectra in Fig.~\ref{ffreq}, the observed deviations between theory and experiment cannot be improved.

Finally, we note that the mechanism of the polarization plane rotation by the gyrotropic birefringence ($c_2$ term in Eq.\,(\ref{chiyy})) is unusual. As shown in the theoretical Section, in the presence of this term the solutions of the Maxwell equations are two linearly polarized modes with polarization planes rotated away from the crystallographic axes $a$ and $b$ (see Fig.\,\ref{fig_th}). Similar to the case of an anisotropic crystal, the light polarized along the $a$-- or $b$--axis is split into ordinary and extraordinary waves, which results in the polarization rotation and ellipticity of the output radiation. In a further simplification, the new optical axes are oriented by exactly $\pm 45^{\circ}$ with respect to the crystallographic axes. This effect is then equivalent to Jones birefringence~\cite{jones_josa_1948, roth_prl_2000}.

\section{Conclusions}

In conclusion, we investigated the rotation of the polarisation plane in magnetoelectric \Yb~under the viewpoint of time and space inversion symmetry arguments. We observe the sign change of the rotation sense under either time or space reversal. This investigation rigorously proves that the polarization rotation in \Yb~must be classified as gyrotropic birefringence. The diagonal terms in the magnetoelectric susceptibility are responsible for the observed rotation due to gyrotropic birefringence. A substantial contribution of the natural optical activity to the polarization rotation could be observed as well. We demonstrate that the latter effect originates from the dynamic magnetoelectric susceptibility.

\subsection*{Acknowledgments}
We thank M. V. Popova for providing us the single crystals of \Yb. This work was supported by the Russian Science Foundation
(16-12-10531) and
the Austrian Science Funds (W1243, I 2816-N27, I 1648-N27).

\end{document}